# Eternalism and Perspectival Realism About the 'Now'

**Matias Slavov[1]**

## Abstract

Eternalism is the view that all times are equally real. The relativity of simultaneity in special relativity backs this up. There is no cosmically extended, self-existing 'now.' This leads to a tricky problem. What makes statements about the present true? I shall approach the problem along the lines of perspectival realism and argue that the choice of the perspective does. To corroborate this point, the Lorentz transformations of special relativity are compared to the structurally similar equations of the Doppler effect. The 'now' is perspectivally real in the same way as a particular electromagnetic spectrum frequency. I also argue that the ontology of time licensed by perspectival realism is more credible in this context than its current alternative, the fragmentalist interpretation of special relativity.

## 1 Introduction

Relativistic physics indicates that the temporal order of two spacelike separated events is relative. Hence there is no absolute present moment. This is evident by contrasting Minkowski spacetime with a classical spacetime. In the supposedly absolute classical spacetime, with an objective unique foliation into simultaneity slices, all and only other events that lie on the same simultaneity slice as the token count as happening 'now.' In relativity, there is no such objective unique foliation. The past, the present, and the future are perspectival in nature. This is good news for eternalism: all temporal locations are equally real.

✉ Matias Slavov
  matias.slavov@tuni.fi

1  Philosophy, Tampere University, Tampere, Finland

As relativity debunks absolute and universal present, it is typically interpreted to favor eternalism over presentism. The most recent challenge of this idea comes from Rovelli [28]. He agrees that modern physics destroys presentism.[1] The structure of relativistic spacetime does not permit us to delineate any privileged extended present. Yet Rovelli does not support eternalism, either. He criticizes Putnam's ([27], Fig. 1 on p. 241) classical argument which employs 'now'-slices. Thus Rovelli ([28], p. 1328) puts it as follows:

> Putnam misinterprets Einstein's simultaneity and mixes relativistic and non relativistic concepts, making up a mess. In particular, Einstein's simultaneity is not a discovery of a fact of the matter about multiple simultaneity surfaces: it is the discovery that simultaneity has no ontological meaning beyond convention. This destroy Presentism, but does not force us into Putnam's Eternalism.

Rovelli also maintains, which is an integral part of his objection, that eternalism may not incorporate passage of time. Here I shall not focus on time's passing, as it is a huge topic of its own. This paper relies on a version of eternalism which is not dependent on hypersurfaces of simultaneity that putatively span across the whole universe. Think the relativity of simultaneity. Provided that the one-way speed of light is invariant,[2] an observer in one frame can conclude that two spatially separated events are simultaneous. Simultaneity within a specific frame is different from mere apparent simultaneity. The observer may calculate that the events which produce their observations are simultaneous (again, based on light's determinate finite speed).[3] In another frame, moving to opposite direction or situated in a different gravitational potential, the very same events are not simultaneous, but successive. The two observers in this scenario have different answers as to what happens 'now.' Both are right in their own frames of reference. This already renders A-locations relative: whether something occurs in the past, in the present, or in the future, is dependent on the perspective. Everybody is right in their own frame of reference. We all have our own moments of 'now.' This is enough for eternalism.

Despite being hospitable to modern physics, eternalism is problematic for truthmaking. What makes statements about temporal positions, like the present moment, true? Armstrong ([1], Chap. 11) defends eternalism because in his view it does not face the typical objections of presentism, like postulating truthmakers for past and future in the present, finding truthmakers outside of time, or accepting non-existences as truthmakers. Armstrong criticizes one option in the debate, but he does not provide a positive answer to the problem. Considering the ramifications of relativistic physics, it is difficult to tell what the truthmakers for the 'now' could be. Statements about what happen 'now' are not true in the paradigmatic sense of truth: correspondence between a sentence and the world. Compare this to a common sensical realism. The table in front of me exists whether I observe, think, or wish it to be there. The substantial existence of the table makes the sentence 'the table exists' true. What renders statements referring to the 'now' true?

---

[1] In his *Order of Time*, Rovelli ([29], p. 44) makes a similar point: "The idea that a well-defined *now* exists throughout the universe is an illusion, an illegitimate extrapolation of our own experience.".

[2] This standard assumption is of course challenged by the conventionality argument. If we use light signals to synchronize two spatially separated clocks, we may only conclude that their roundtrip time is invariant. Light's isotropic speed is not directly measurable. Consequently, spatially distant events are not factually but only conventionally simultaneous. This idea was originally formulated by Einstein and Poincaré, subsequently developed by Reichenbach and Grünbaum, and then criticized by Malament. Thyssen [35] details these stances and how they are related.

[3] For the distinction between real and apparent simultaneity within an inertial frame, see Slavov ([31], p. 348).

I shall approach this problem along the lines of perspectival realism. This doctrine has recently been established by Massimi [21] and her research group in the philosophy of science. The novelty of this paper is to apply the idea of perspectival realism in the philosophy of time. This means that although physical events are taken to be primitive, mind-independent existences, temporal locations like the 'now' are dependent on a perspective. There is no truthmaker out there to be discovered that truthmakes claims concerning the present moment. Instead, choosing the perspective makes statements referring to the present true.

To that end, the rest of the paper proceeds as follows. The next section advances an argument in which perspectives play a fundamental role in the existence of the present time. To corroborate the argument, the 'nows' of the observers shall be assimilated with electromagnetic wave lengths, as the Lorentz equations in relativity are analogous to those of the Doppler effect. The subsequent section explains why perspectival realism provides a more plausible ontology of time than its relevant alternative, the fragmentalist interpretation of special relativity. The closing section ends by claiming that one can convincingly incorporate perspectival realism about the present moment in eternalist metaphysics.

## 2 Choosing the Perspective

It is well-known that for the presentist, finding truthmakers about the past or future is problematic. If the past is no longer existent, and the future does not yet exist, there is nothing that makes statements concerning them true. In the view of no-futurism, truth-claims concerning the future are troublesome as there is a fixed past and open future. The eternalist does not face a similar kind of problem, because all temporal positions exist evenly.

Despite such initial optimism—or the provisional assumption that the eternalist is better off with truthmaking than the presentist[4]—finding truthmakers for eternalism is all but simple. At first sight, we might think that events could serve as truthmakers. They are the fundamental entities of relativity. Their existence, unlike their temporal order, is in every case substantial. But events themselves are not moments. Without choosing some perspective, an event is not associated with any temporal position. The existence of events is not a perspectival matter, but their space-like temporal order is. To show that events exist independently, but not moments, consider the spacetime diagrams below. Imaged are three events, A, B, and C.[5] The white rounds denote the 'now' of an event, or events, while the black rounds indicate future events in spacetime from an observer's point of view (Figs. 1, 2, 3).

---

[4] Baron [4], Cameron [8] and Tallant [34], among others, have explored truthmaking in the presentist context.

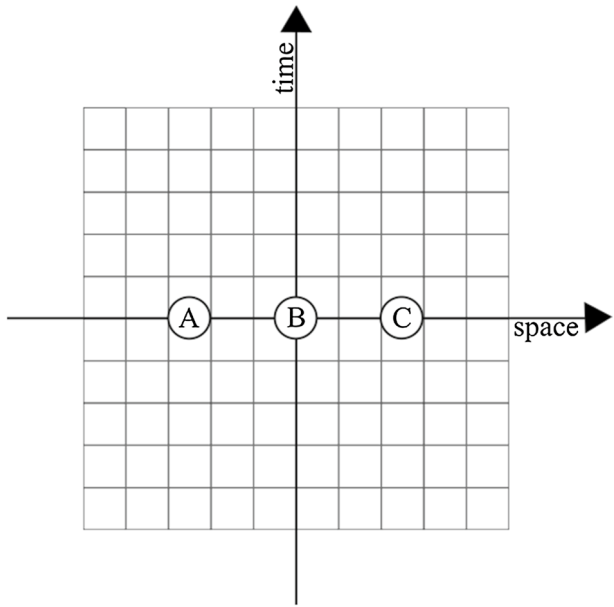

**Fig. 1** For an observer with zero velocity, the three events are simultaneous. They happen 'now'

All observers disagree on what happens 'now,' but no one disagrees whether the events happen or not [9], p. 179). In accordance with special relativity, events exist in their own right. One way to explain this is to note that observers do not generally observe the events at the exact same time in which they happen in a specific frame. This is due to time-lag: the information of something happening is transferred by light (or some other electromagnetic spectrum frequency or a signal slower than it) from the event to the observer. The event happens before and independently of it being observed.

So, events themselves are not relative—they just happen—but the temporal order of spacelike separated events is. Accordingly, events, however important for relativity's ontology, *are not* the truthmakers of claims about 'now.' This is a truly stunning implication for our metaphysics. It is like the 'now' vanishes; as if it is not affiliated with anything real. Yet I think there is a way to argue for a type of realist ontology of the 'now.' Below I present a mathematical analogy that explains how specifying a particular frame makes statements concerning the present true.

---

[5] I have drawn these images by adopting the figure in Bardon ([2], p. 91), rendered in digital form by Marcia Underwood. The original file is the courtesy of Wikipedia user Acdx, based on file: Relativity_ of_Simultaneity.svg. Source (consulted on September 22, 2020): https://en.wikipedia.org/wiki/Relativity _of_simultaneity#/media/File:Relativity_of_Simultaneity_Animation.gif. The original file includes an animated 'now'-slice, and it shows the skewing of spacetime according to different observers. I provide these figures to highlight the independence of events, and their separation of moments. It should be made clear that there are various ways of conceiving spacetime. Gilmore et al. [16] elucidate three different positions on the four-dimensional views of the universe implied by relativity theory: (1) spacetime unitism, (2) the B-theory, and (3) perdurantism.

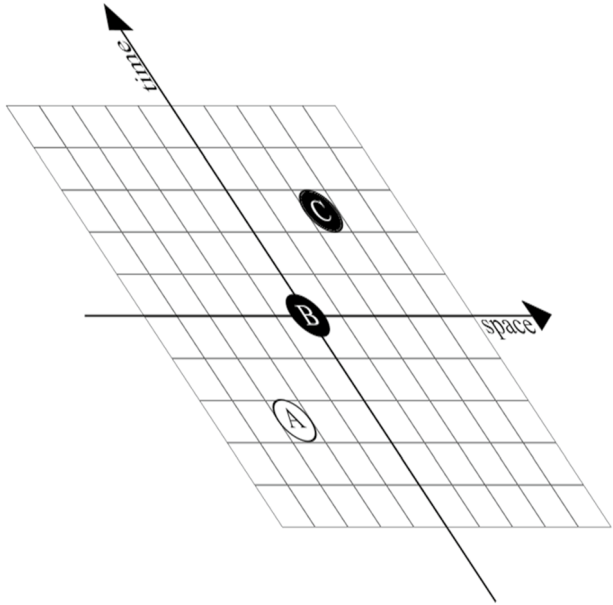

**Fig. 2** For an observer moving with *v*, A happens 'now'

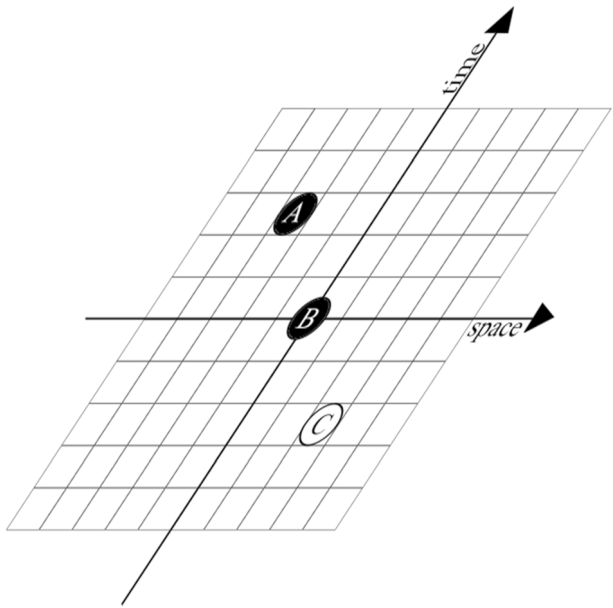

**Fig. 3** For an observer moving with −*v*, C happens 'now'

If for $O'$ two events are simultaneous, $\Delta t' = 0$. From Lorentz transformations we get that for $O$, who moves with a velocity $v$, the time difference between the events is $\Delta t = v\Delta x'/c^2 \sqrt{1 - \beta^2}$, in which $\beta = v/c$ and $\Delta x' \neq 0$. The two distant events are successive for $O$. For a third observer $O''$ moving to opposite direction with a velocity $-v$, the events would be successive, and could be in the opposite order than for $O$. Whether two events are simultaneous or not depends on the selection of the inertial frame of reference. Consequently, the three observers all have different 'now' moments.

Now to the central allegory of this paper. The relativity of temporal positions is analogous to the relativity of electromagnetic waving. Here is an example of the Doppler effect. If a stationary source emits waves with a wavelength of $\lambda_0$, then the waves for an approaching source are shorter and higher in frequency, $\lambda_+ = \sqrt{1 - \beta/1 + \beta}\lambda_0$, and for a receding source they are longer and lower in frequency, $\lambda_- = \sqrt{1 + \beta/1 - \beta}\lambda_0$. All the values for the same electromagnetic wave, $\lambda_0, \lambda_+$, and $\lambda_-$ are different.

My argument does not assume that waving and temporal positions are perfectly equivalent. Color perception, for example, differs in many ways from temporal experience. But the two following questions—What determines the electromagnetic frequency? What determines the temporal order of events?—Have the same answer: the selected frame of reference. We can speak of a definite frequency or a definite order only after designating the observer. In both cases, the relevant quantities are relative to motion (and by general relativistic extension, both quantities are relative to gravitational effects).

Both electromagnetic waving and physical events exist by themselves.[6] But without specifying a frame, there is no sense asking what is the one true spectrum frequency, any more than there is sense asking what really happens right now. In this respect, redshifting/blueshifting in cosmology is analogous with the relativity of event ordering in special relativity. Depending on the co-ordinate system, one observer might observe *the same electromagnetic waving* as a color red, the other as a color green, and the third as a color blue. Likewise, depending on the co-ordinate system, observers might observe *the same events* in different order, for example, one sees the opening of a living room door first and opening of the kitchen window after, the second both openings at the same time, and the third the opening of the window before the door. Waves and events themselves do not dictate frequencies or temporal positions; the choice of the frame does. In the same way as my perception of some color is real to me, also the happening of an event 'now' is real to me. Such reality must be perspectival in nature.

---

[6] Electromagnetic waves are made of real entities, electric and magnetic fields. They are not mere theoretical instruments: they restore energy and carry information. Likewise, physical events just happen absolutely before anyone sees them [because an event happening and an observer registering it are timelike (or lightlike) separated: the information from the earlier event (in this case, the event itself) reaches the later event (observer's perception) with the electromagnetic wave speed or slower than it], and so must exist independently of the observer.

It might be argued that the analogy between the relativity of temporal positions and electromagnetic frequencies is not completely straightforward. In some sense there is a preferred reference frame for electromagnetic frequency: the inertial frame in which the emitter is momentarily at rest. This proper frequency could be equated with the true frequency. Events are not associated with any preferred inertial frames because they, treated as point-like, have zero duration. This is different from electromagnetic signals as their emission occurs during a nonzero interval. A part of the observer's lightlike wordline is bound by the emitter. This is consistent with infinitesimal inertial reference frames in general relativity, too.

This counterargument lacks substance. First, concerning the debates within the ontology of time, it is controversial to treat events merely as point-like. Ben-Yami [6] has convincingly argued that real applicable events have durations. Durationless mathematical points are not physical events. They are idealizations which take "a process of precisification to its mathematical, theoretical limit" [6], p. 1358). The duration of an actual event is also rather vague. The beginning and the end of an event cannot be defined with a high degree of accuracy. Neither of the event's temporal boundaries is clear-cut. Second, it is true that, originally in the emitter's inertial frame, one can define a proper frequency. Certain radio antenna, for example, has charges that oscillate 100 million times/s so to produce a frequency of 100 MHz. Definite electric currents are needed at the source, but not afterwards. Once the wave has been created and it has left its original source, the received frequency becomes a matter of perspective. The result is that there is no one true frequency but a multiplicity of true frequencies. This reinforces, rather than questions, the perspectival nature of frequency.

An important characteristic in the notion of perspectival reality is that there is truth within a perspective. Giere ([15], p. 81) elucidates this position in an indexical fashion. In his example, an observer standing on the steps of the Capitol Building sees the White House on the right side of the Washington Monument. This statement is true within the chosen perspective. It would be false to say that, for example, the White House is on the left of the Washington Monument. Once the perspective is explicated, there is little point to argue about the veracity of a conclusion like this. Giere ([15], pp. 81–82) notes that

> Of course, we do not typically think of our everyday ways of speaking as being within a perspective. But neither do most people think of colors as being relative to the perspective of the human chromatic visual system.

There are truths about spatial locations as well as color perceptions within a perspective. Correspondingly, there are truths about statements concerning the 'now' within a perspective. In our everyday lives, we do not think that tensed claims are made from a perspective. But this is no different than our everyday claims that tacitly assume objective and absolute character of colors. We tend to say that grass is green, assuming that greenness is the property of the object. Still color perception is relative to state of motion and sensory and nervous systems. As long as we specify the circumstances under which particular statements are true, there is no reason to be skeptical or cynical about them. 'Grass is green' is no doubt true, but it is true only within a perspective.

Recently, in his defense of the B-theory, Boccardi [7] has claimed that A-determinations, like presentness, are irreal. The main argument in his paper is that B-theoretical approach can explain our dynamic experience of time better than the A-theory. In the course of articulating his stance, Boccardi ([7], p. 395) writes:

> A-determinations of pastness, presentness and futurity are taken to be unreal and illusory. According to the B-theory, objective A-determinations are *totally* absent from reality, even from that part of reality which consists of experiential events.

Boccardi's metaphysical position is that A-properties are not fundamental or ultimate features of the world. The eternalist and relativistic framework assumed in this paper agrees that reality is not inherently tensed. But why is this different than saying that reality is not inherently colored? Non-fundamental A-properties may be included in the list of what exists in the same way as non-fundamental colors.[7] Here we may draw on the notion of relative fundamentality, as articulated for example by Tahko [33]. Less fundamental entities depend on the existence of more fundamental entities in terms of priority ordering. This implies that the more fundamental may exist independently of the less fundamental, but not the other way around. Yet relative fundamentality does not imply that the less fundamental is somehow less real or non-existent.

In no way does relativity imply that temporal ordering is fictitious. Rather, what is imaginary is that there would be an absolute-universal time, which functions as an objective metric for placing events into past, present or future. Contradicting truth-claims about the present moment can be resolved once a frame is specified,[8] and the notion of a perspectival reality adopted.

## 3 Tackling the Fragmentalist Interpretation of Special Relativity

Perspectival realism licenses a different kind of ontology compared to its relevant alternative, the fragmentalist interpretation of special relativity. According to Lipman's ([19], p. 23) general definition, which draws on the work of Fine [14], "fragmentalism is the view that the world is inherently perspectival." Fragmentalism does not maintain only that we have different perspectives and ways of representing the world. Reality itself is thought to be essentially perspectival, constituted by fragmented facts. Some aspects of the fragmentalist position are not in tension with the version of perspectival realism as defended in this paper. A-properties are inherently perspectival and real. Different temporal locations are not "*mere appearances of an underlying layer of compatible facts,*" to paraphrase Lipman ([19], p. 24). What varies across perspectives like reference frames is not mere appearance.[9] This much is consistent with fragmentalism. It is however unclear whether fragmentalism separates events and their temporal locations. According to perspectival realism, events, on the side of the world, form the material basis for our tensed claims. Unlike frames of reference, they are fundamental.[10] Events occur as they do, independently of observers. Their existence is not perspectival. In my rendition, fragmentalism does not take events to be primitive. Instead, in the context of special relativity, fragmental frame-times are fundamental. Consider the following general characterization by Lipman ([19], p. 23):

> We standardly assume that we only ever have perspectival representations of a non-perspectival world. Fragmentalism denies this assumption, allowing that the world is itself an inherently perspectival place where facts do not simply obtain or fail to obtain, as we ordinarily assume, but where certain facts can obtain in the context of one set of facts and yet fail to obtain in the context of other sets of facts.

---

[7] Slavov [31, 32] has argued that A-properties, although less fundamental than B-properties, may be incorporated in non-reductive physicalism.

[8] Pinillos [24] clarifies the contextual nature of truth regarding time dilation. In his example, there is a passenger on a train and an observer at a platform. The observer stays at the platform and the train makes a roundtrip. Eventually the two compare their clocks and conclude that they do not agree. They have measured different lapses of time between two events, that is, between the departure and the arrival of the train. When asking whether the passenger or the observer is right, Pinillos ([24], p. 65) notes that there are three possible answers: "(A) Exactly one of them asserted something true, (B) Neither one of them asserted anything true, and (C) Both asserted something true." The last option is true without any contradiction.

In the quote above, fragmentalism is associated with denying the assumption that we "have perspectival representations of a non-perspectival world." This is the major difference between fragmentalism and perspectival realism. The latter treats A-locations as "perspectival representations" of "non-perspectival world" of physical events. To continue with the next sentence of the quote, Lipman refers to the concept of a fact. He does not elaborate on this notion. Provided that events are treated factually, there should be an unambiguous fact to the matter of whether an event obtains. As Dieks ([11], p. 170) has noted, the four-dimensional relativistic diagram records all events spread across spacetime.[11] In no case may events fail to obtain. If an event does not obtain, it does not exist. It is not then a part and parcel of spacetime. A non-existent thing is not an event, and never will be. Events have to be fundamental. They are not perspectival or separated into different fragments. The claim that events exist independently of them being observed is perfectly common sensical and consistent with relativity. Due to time-lag, physical events take place before we receive information of them.[12] If we do not affect the events in question, they must exist substantially.

---

[9] In his characterization of the metaphysical consequences of special relativity, Lipman ([19], p. 22) takes the following to be the standard view: "The Minkowskian conception of spacetime accepts the assumption that what varies across perspectives (such as frames of references) must be mere appearance." This is a controversial characterization. Agreeing with standard relativistic physics, Shimony ([30], Chap. 18), by leaning on the contributions of Carnap, Grünbaum and Weyl, has argued in detail that frame-dependent A-properties are subjective but still in no way "mere appearances.".

[10] In its inception, special relativity relied heavily on inertial frames (a good collection of early papers is Lorentz et al. [20]). Even later expositions, like those of Prokhovnik [26] and Bell [5], were sympathetic to the view that there is a frame really at rest, defined by the aether. These are certainly interesting historical considerations. Yet more recently Maudlin ([22], p. 67) has criticized the primacy of inertial frames by arguing that they are not fundamental, but derivative. The notion of an inertial frame refers to motion. No particle of mass or body has absolute speed. Moreover, there are no inertial frames in general relativity.

[11] Whether events should be treated statically or dynamically (so to include passage into one's ontology) is a debate of its own. Eddington ([13], p. 46) famously provided a static analysis: "Events do not happen; they are just there, and we come across them." Dieks (2006) objects this and argues for a dynamic nature of events that is consistent with the block universe view.

In themselves events are however entirely mute on whether a moment is past, present, or future. In other words, events are necessary for temporal placement (A-properties of past, present, and future) and temporal ordering (B-properties of before and after), but they do not necessitate the truths about any moment. Truthmaking is done only by choosing the perspective. This can be either an inertial frame as in the special theory or a non-inertial like in the general theory. It is a conventional matter of how many co-ordinate systems one wishes to establish. There is no privileged frame, but there are as many times as there are frames [10], p. 195). A-properties are not the attributes of physical processes. The choice of the frame determines an event's temporal, tensed location: 'nowness' of an event is an observer-dependent matter.

The fragmentalist position brings back partly Newtonian ideas by invoking "genuine facts regarding absolute simultaneity, duration and length" [19], p. 21). Fine supports this by arguing that reality is fragmented into parts that are all affiliated with their own frame-times. Each fragment is considered Newtonian in a way that saves an absolute notion of the present. In Fine's ([14], p. 304) own words:

> The presentist takes there to be an absolute and objective sense in which a given frame-time is the standpoint of reality. He is therefore in a position to distinguish a particular frame as the frame of this standpoint; and this then enables him to characterize an absolute notion of simultaneity as simultaneity within this frame.

Based on the quote above, any given inertial frame can be chosen as an absolute frame having its own notion of simultaneity. The terminology at play is not explicit. I assume that in this context 'absolute' does not denote 'unique'. Fragmentalism does not imply that there is one and only one way to foliate spacetime. Rather, different fragments of reality incorporate their own genuine tensed realities. "This then," to reiterate, enables one "to characterize an absolute notion of simultaneity as simultaneity within this frame."

How tenable this view is depends on the notion of 'absolute.' According to a standard definition, at least in the philosophy of time context, 'absolute' means substantial existence. When Newton [23] described absolute time in the Scholium of the Definitions of his *Principia*, he very clearly meant that time "in and of itself and of its own nature, without reference to anything external, flows uniformly." He separated the substantial, self-existing (that is, absolute) time from observable and relative measures of time. Fragmentalism is certainly not committed to the substantial Newtonian flow of time. But it still supports a qualified substantivalist claim:

---

[12] On the compatibility of time-lag and eternalism, see Power [25].

fragments of reality are *by themselves* associated with absolute simultaneity. Locating the present moment is not about choosing a perspective, about setting up a frame, but identifying a fragment's own absolute time.

It is hard to see how fragmentalism could completely avoid obsolete Newtonian ideas. If fragments of reality have their absolute frame-times, one is committed to local absolute spaces. As with time, fragmentalism is not implying that there is one and only one space. Neither does it claim that absolute space is immaterial as Newton had it. Yet fragmentalism updates Newton's substantivalist concept of time (and space) with a multiplicity of substantivalist frame-times. This is at odds with relativity. Hofweber and Lange [18] detail how the fragmentalist take on relativity is in tension with the Lorentz transformations. This includes, for example, the relations of salient variables like spacetime coordinates in different frames, as well as other physical facts such as specific electric fields, pressures and charge densities.

Whether fragmentalism could be made consistent with modern science has to do with how well it can be integrated to various physical theories. Lipman ([19], p. 22) acknowledges the tentative nature of his research program. In his view, "the question of which theory is ultimately better will depend crucially on wider theory integration, how well each interpretation generalizes and meshes with other scientific theories". I agree that this is a difficult topic, and not to be settled in any easy way. To my understanding, the Newtonian aspects of fragmentalism are in tension with the consensus of institutionalized physics community. The special theory essentially shows that time is relative, not absolute. The theory does not exist in isolation, because it is a central part of modern physics. The special theory corrected the preceding Maxwellian–Hertzian ether-based electrodynamic theories, and it is part of experimental high-energy physics, even the quantum field theory. It is also contained in general relativity which accounts for gravitational fields [3], p. 9).

Here I am, perhaps in a controversial manner, appealing to the authority of a current scientific consensus at the expense of a metaphysical theory. It is certainly debatable how naturalized our philosophy of time ought to be.[13] Leaving aside important metametaphysical debates, I wish to stress the following point. Fragmentalism maintains the absoluteness of the present moment in the hybrid Newtonian-relativistic sense. A problem emerges: the relative quantity of time fails to be comparable to the relative quantity of electromagnetic spectrum frequency. This is highly dubious; the equations of the Doppler effect and special relativity are structurally analogous. They both show how specific quantities are transformed from one frame to another. If we allow the present moment to be absolute, why not allow colors to be absolute, too? Say we are having lunch here on Earth. There is no absolute fact to the matter of whether the Mars rover takes a picture at the exact same time as we are eating our meals. Fragmentalism maintains that these different fragments, the specific inertial frame-time fragment on Earth, and the specific inertial frame-time fragment on Mars, both have their own absolute simultaneities and hence present times. If this is true, why would they not have their own absolute colors? If 'nowenesses' are properties of different fragments, why are colors not properties of different fragments? A-determinations and colors do not have a substantial existence. Colors are not absolute attributes of any fragments of reality. Neither are present moments.

---

[13] For discussion, see Hawley [17] and Dyke [12].

Fine and Lipman could acknowledge that their interpretation upends the standardly assumed order of relative fundamentality regarding spacetime and inertial frames. They could insist that this is not a mark against their fragmentalism because it is still an empirically equivalent interpretation to the orthodox theory. They could also maintain that their interpretation does not satisfy my Doppler effect analogy, but this metaphor is hardly essential to understanding time in relativity theory. Within the relativistic context, this would be however cherry picking: one cannot consistently choose what quantities to hold absolute and what relative. Relative simultaneity is encoded in the Lorentz transformations. The equations show that different frames may not agree on the temporal interval between two events. As with color quantities dictated by the Doppler effect, this is usually true on a cosmic scale. Any present moment is inherently relative and subjective much the same way any color is inherently relative and subjective. This is not only a claim about the mathematical structure of a scientific theory. Lorentz equations tell us in what way temporal locations exist. The ontology set forth by perspectival realism fits better with these equations, and hence with the contemporary scientific practice.

## 4 Conclusion

This paper was directed at the question of what, in an eternalist relativistic account of the world, makes statements concerning the present true. More broadly, it tackled the ontological issue: In what sense does the 'now' exist? It was argued that the 'now' does not exist like ordinary physical objects or processes like events. Ascertaining what makes statements concerning the present moment true is different than ascertaining what makes commonplace statements like 'there are stars in the sky' true. The 'now' is not out there to be discovered like astronomical objects.

As the 'now' lacks a truthmaker in mind-independent physical reality, this paper advanced the argument that, under the assumption of perspectival reality, the choice of the perspective renders claims concerning the present true. The existence of A-properties is relative in the same indexical way as the existence of spatial locations and electromagnetic spectrum frequencies. This argument is bolstered by the fact that the Lorentz transformations are similar in structure to the equations of the Doppler effect. In conclusion, a perspectival realist position about the present moment is consistent with eternalist metaphysics and relativistic physics.